\documentclass[12pt]{iopart}
\usepackage{iopams}

\usepackage{graphicx}
\usepackage{dcolumn}
\usepackage{bm}
\usepackage[utf8]{inputenc}
\usepackage[T1]{fontenc}
\usepackage{mathptmx} 
\usepackage{amsmath}

\usepackage{float}[]
\usepackage{longtable}

\usepackage{subfigure}
\usepackage[export]{adjustbox}
\usepackage{feynmp}

\usepackage{todonotes}

\begin{document}

\title[Random Walk on a Random Surface]{Random Walk on a Random Surface:
Implications of Non-perturbative Concepts and 
Dynamical Emergence of Galilean Symmetry}

\author{N.~V.~Antonov,$^{1,2}$ 
N.~M.~Gulitskiy,$^{1,2}$  
P.~I.~Kakin$^{1*}$ 
and A.~S.~Romanchuk$^{1}$}

\address{$^1$ Department of Physics, St Petersburg State University,
7/9 Universitetskaya Naberezhnaya, St Petersburg 199034, Russia \\
$^2$ N.~N.~Bogoliubov Laboratory of Theoretical Physics, Joint Institute for Nuclear Research, Dubna 141980, Moscow Region, Russia}

\ead{n.antonov@spbu.ru, n.gulitskiy@spbu.ru, p.kakin@spbu.ru, st108315@student.spbu.ru}
\vspace{10pt}

\begin{abstract}
We study a model of random walk on a fluctuating rough surface using the field-theoretic renormalization group (RG).  The surface is modelled by the well-known Kardar--Parisi--Zhang (KPZ) stochastic equation while the random walk is described by the standard diffusion equation for a particle in a uniform gravitational field. In the RG approach, possible types of infrared (IR) asymptotic (long-time, large-distance) behaviour are determined by IR attractive fixed points. Within the one-loop RG calculation (the leading order in $\varepsilon=2-d$, $d$ being the spatial dimension),
we found six possible fixed points or curves of points. Two of them can be IR attractive: the Gaussian point (free theory) and the nontrivial point where the KPZ surface is rough but its interaction with the random walk is irrelevant. For those perturbative fixed points, the spreading law for a particles' cloud coincides with that for ordinary random walk, $R(t)\sim t^{1/2}$. 
We also explored consequences of the presumed existence in the KPZ model of a non-perturbative strong-coupling fixed point. We found that it gives rise to an IR attractive fixed point in the full-scale model with the nontrivial spreading law $R(t)\sim t^{1/z}$, where the exponent $z<2$ can be inferred from the non-perturbative analysis of the KPZ model.
Thus, the spreading becomes faster on a rough fluctuating surface in comparison to a smooth one. 
What is more, the Galilean-type symmetry inherent for the pure KPZ model 
extends dynamically to the IR asymptotic behaviour of the Green's functions of the full model.

*Author to whom any correspondence should be addressed.
\end{abstract}

%\pacs{05.10.Cc, 05.70.Fh}

\section{\label{sec:level1} Introduction}

Over many decades, constant attention was attracted to random walks, diffusion processes and particles transfer in complex environments such as random, disordered, inhomogeneous, porous or turbulent media.
They demonstrate interesting behaviour: sub- and superdiffusion, anomalous (non-Brownian) spreading laws, relation to the origin of the $1/f$ noise, trapping, etc.; see, e.g.~\cite{MP}~-- \cite{Walks4} and references therein.

Another long-standing vast area of research is devoted to 
stochastic growth processes, kinetic roughening phenomena and fluctuating surfaces or interfaces. The most prominent examples include deposition of a substance on a surface and the growth of the corresponding phase boundary; propagation of smoke, fire and solidification fronts; erosion of landscapes and seabed profiles; growth of bacterial colonies, tumors and vicinal surfaces, molecular beam epitaxy and many others; see, e.g.~\cite{Eden}~-- \cite{Xia} and references therein.

In the previous work~\cite{Dima}, we combined these two phenomena
and studied random walk on a random rough surface, modelled by the generalized Gaussian stochastic Edwards--Wilkinson ensemble~\cite{EW}. The analysis, based on the field theoretic renormalization group (RG), has shown that the interplay between the two random processes leads to nontrivial rich behaviour: the~Green's functions of the full model
demonstrate various types of self-similar (scaling) asymptotic forms in the infrared (IR) range (large scales and long times),  while the spreading of a cloud of particles obeys power laws different from those for ordinary random walk.
The stability regions for those asymptotic regimes were identified
and the corresponding critical exponents were determined exactly.

Thus, it is interesting to consider more realistic models for the fluctuating surfaces and to explore how the results can depend on their statistics.
In this paper, we study another model of random walk on a rough fluctuating surface. We consider the diffusion equation for a particle in a uniform gravitational field. The surface is modelled by the Kardar--Parisi--Zhang (KPZ) stochastic equation for the height field~\cite{KPZ}.\footnote{It is worth noting that a model equivalent to~\cite{KPZ} was introduced earlier by Forster, Nelson and Stephen~\cite{FNS1} as a $d$-dimensional stochastic generalization of the Burgers equation in terms of the potential vector field.} 

In section~\ref{Model}, we give detailed description of the full stochastic model and 
provide its field theoretic formulation. The original KPZ model possesses a certain symmetry with respect to a small tilt~\cite{KPZ} which in the vector formulation takes on the form of Galilean invariance~\cite{FNS1}. In the full-scale model that involves the diffusion equation, this symmetry is in general violated: it survives only for a special choice of the coupling constants, not justified by their physical origin. In section~\ref{QFT}, we present canonical dimensions of the fields and the parameters, analyze the form of the needed counterterms and establish multiplicative renormalizability of the full model. It appears logarithmic (all the couplings become dimensionless) for the spatial dimension  $d=2$, so that the deviation from logarithmicity  $\varepsilon=2-d$ plays the role of the formal small parameter in the RG analysis. The renormalization constants are presented in the leading-order (one-loop) approximation; the diagrammatic techniques and details of the calculation are included
in the Appendix.

In section \ref{RGEq}, we present the corresponding RG equation and provide the leading-order expressions for the RG functions (anomalous dimensions $\gamma$ and $\beta$ functions).
In a renormalizable model, possible IR asymptotic scaling regimes (long times, large distances) are determined by IR fixed points of the RG equations. Within the one-loop approximation, 
we established existence of six fixed points (or rather curves of points). Only two of them can be IR attractive: the Gaussian point (free theory) and the nontrivial point where the KPZ surface is rough but its interaction with the random walk is irrelevant. 
The corresponding critical dimensions are presented in  section~\ref{Scaling}.
Although the calculation was performed only in the leading approximation
(first order of the expansion in $\varepsilon$), these expressions appear to be exact (no corrections of the order $\varepsilon^2$ and higher). For the both points, the spreading law for a particles' cloud coincides with that for ordinary random walk, $R(t)\sim t^{1/2}$. 

However, the situation is not quite satisfactory. The coordinates of the non-trivial attractive fixed point lie in the unphysical area of parameters which is the well-known flaw of the perturbative RG analysis of the KPZ model; see, e.g.~\cite{Wiese2}~-- \cite{Wiese}. 
Fortunately, various non-perturbative considerations (numerical simulations, mode-coupling theories, functional RG) suggest the 
existence of a certain strong-coupling IR attractive fixed point, undetectable in any
perturbative scheme; see, e.g.~\cite{Kinzel}~-- \cite{Canet4}.

In  section~\ref{KPZ-NP}, we investigate the consequences of the presumed existence of such hypothetical point onto the IR asymptotic behaviour of our full-scale model.
We show that it gives rise to a non-trivial IR attractive fixed point in the physical range of parameters of the full model. As a special illustrative example we show in section~\ref{Spread},
that this leads to superdiffusive spreading law $R(t)\sim t^{1/z}$ for the mean-square radius of the particles' cloud, where the exponent $z<2$ can be inferred from the non-perturbative analysis of the KPZ model~\cite{Gomes}~-- \cite{Ala1}.
Thus, the spreading becomes faster on a rough fluctuating surface in comparison to a smooth one. 
What is more, the Galilean-type symmetry inherent for the pure KPZ model and violated by the
inclusion of the diffusion equation for the particles' density is 
dynamically restored for the IR asymptotic behaviour of the Green's functions of the full model.
   
Section~\ref{Conc} is reserved for discussion and conclusion.

\section{Description of the model. Field theoretic formulation}
\label{Model}

A point particle randomly walks over a $d$-dimensional surface that randomly grows in its turn; the latter is embedded into the $(d+1)$-dimensional space. 
The particle's location is given by the surface height $h(t,{\bf x})$, where ${\bf x}(t) =\{x_i(t)\}$ is projection of the particle's coordinates on the  $d$-dimensional substrate, $i=1\dots d$.

Such a setup does not allow for the particle to leave the surface by either jumping or falling off; indeed, coordinate ${x_i}$ with $i=(d+1)$ is restricted to the surface, $h(t,{\bf x})\equiv {x_{(d+1)}}$, and does not play a role of independent Cartesian coordinate. This is  a simplification that we use to construct the model.

The basic stochastic equation of motion for a particle located at the point ${\bf x}(t) =\{x_i(t)\}$ in an external drift field $F$ is~\cite{MP}~-- \cite{Walks3}:
\begin{equation}
\partial_t x_i = F_i(t,{\bf x}) + \zeta_i, \quad 
\langle \zeta_i(t)\zeta_j(t') \rangle =2\kappa_0\delta(t-t').
\label{RW}
\end{equation}
Here $\zeta_i=\zeta_i(t)$ is a Gaussian noise with zero mean and the given pair correlation function, $\kappa_0>0$ is the diffusion coefficient, and $\partial_t=\partial/\partial t$ is derivative over time coordinate $t$.  
Here and below, the subscript ``$0$'' refers to bare parameters and is used to differentiate them from the renormalized ones.

In the macroscopic description, the particles' density $\theta(t,{\bf x})$ replaces location ${\bf x}(t)$ and satisfies the Fokker-Planck equation that for the model~(\ref{RW})
can be reduced to the diffusion-type equation (see, e.g. 
section~5.7 in~\cite{Vasiliev}):
\begin{equation}
\left\{ \partial_t + \partial_i (F_i - \kappa_0\partial_i)\right\}\, \theta(t,{\bf x})=0;
\label{FPE}
\end{equation}
here and below, $\partial_i=\partial/\partial {x_i}$ and the summation over repeated indices is implied.

When choosing the form of the drift $F$ in a gravitational field, its symmetries should be taken into account: the $O(d)$ symmetry and the invariance with the respect to the transformation $h\to h+$const. Gradients of the field $h$ satisfy these conditions:
\begin{equation}
F_i=  -\partial_i Y(h) \simeq  - \alpha_0 \partial_i h + \dots,
\label{gravi}
\end{equation}
with a certain function $Y(h)$ whose higher-order terms of the expansion
in $h$ and $\partial$ can be dropped as being IR irrelevant (in the sense of Wilson), i.e., not contributing to the behaviour in the IR range.
The factor $\alpha_0>0$ is roughly proportional to the particle's mass and the gravitational acceleration. 

In our preceding study~\cite{Dima}, the height field $h(x)$ was modelled by the (generalized) Edwards--Wilkinson \textit{linear} stochastic differential equation~\cite{EW}. Here, we employ a more realistic (and widely regarded as fundamental) non-linear KPZ equation~\cite{KPZ}:
\begin{equation}
\partial_t h = \sigma_0 \partial^2 h + V(h) + f,  
\label{equation1}    
\end{equation}
where $\sigma_0>0$ is the surface tension coefficient and $\partial^2=\partial_i\partial_i$. The non-linearity $V(h)$ is taken in the form
\begin{equation}
V(h)= \frac{\lambda_0}{2}\, (\partial h)^2 =
\frac{\lambda_0}{2}\,(\partial_i h)\,(\partial_i h),
\label{KPZ}    
\end{equation}
where the coupling constant ${\lambda_0}$ can be of either sign. Furthermore, $f=f(x)$ is a random Gaussian noise with zero mean $\langle f\rangle=0$ and the pair correlation function\footnote{
From equations~(\ref{equation1}), (\ref{KPZ}),
it follows that the mean value $\langle h\rangle$ grows linearly in time,
in agreement with its physical meaning. However, 
one is usually interested in the equal-time correlations around 
this moving interface. Thus, the mean value $\langle f\rangle$ is chosen
to cancel this growth. In practice, both mean values can be ignored simultaneously, which fits well with formal prescriptions of analytic regularization; see, e.g.~\cite{KPZ}.}
\begin{equation}
\langle f(x)f(x')\rangle = D_0\, \delta(x-x') =
D_0\,\delta(t-t')\, \delta^{(d)}({\bf x}-{\bf x}'), \quad D_0>0.
\label{noise}    
\end{equation}

Alternatively, the equation~(\ref{equation1}), (\ref{KPZ}) can be rewritten in the form of the stochastic $d$-dimensional Burgers equation for a purely potential velocity field $v_i=\partial_i h$~\cite{FNS1}:
\begin{equation}
\nabla_t v_i = \sigma_0 \partial^2  v_i  + f_i,
\label{Burgers}    
\end{equation}
with the random force $f_i= \partial_i f$ and the covariant (Lagrangian, material) derivative
\begin{equation}
\nabla_t = \partial_t - \lambda_0  (v_k\partial_k),
\label{Burgers2}    
\end{equation}
where we used the obvious relation $\partial_i v_k = \partial_k v_i$.

Substituting the force~(\ref{gravi}) with the random height field from 
(\ref{equation1})~-- (\ref{noise}) into the diffusion equation~(\ref{FPE}) turns the latter into a stochastic equation:
\begin{equation}
\nabla_t' \theta =  \kappa_0 \partial^2\theta, 
\label{Kolya}
\end{equation}
with its own material derivative
\begin{equation}
\nabla_t' = \partial_t - \alpha_0 \partial_k(v_k \cdot \,\,).
\label{Burgers3}    
\end{equation}

According to the general Martin--Siggia--Rose--De~Dominicis--Janssen  theorem (see, e.g. section~5.3 in monograph~\cite{Vasiliev} and references therein), the full stochastic problem~(\ref{FPE})~-- (\ref{noise}) and~(\ref{Kolya}), (\ref{Burgers3}) 
is equivalent to the field theoretic model for the doubled set of fields $\Phi = \{\theta',h', \theta, h\}$ with the de~Dominicis--Janssen action functional:
\begin{eqnarray}
{\cal S}(\Phi) = {\cal S}_{\theta}(\Phi) + {\cal S}_{h}(h',h),
\label{Action}
\end{eqnarray}
where 
\begin{eqnarray}
{\cal S}_{\theta}(\Phi) = \theta' \left[-\partial_t \theta + \kappa_0 \partial^2\theta + \alpha_0 \partial (\theta \partial h) \right]=
\theta' \left[-\nabla'_t \theta + \kappa_0 \partial^2\theta \right]
\label{ActionT}
\end{eqnarray}
with $\nabla_t'$ from~(\ref{Burgers3}) and
\begin{eqnarray}
{\cal S}_h(h',h) = \frac{1}{2} h'D_0 h'+ h'\left[-\partial_t + \sigma_0 \partial^{2} +\frac{\lambda_0}{2}\, (\partial h)^2 \right] h.
\label{Actionh}
\end{eqnarray}
Here $\theta'$, $h'$ are the Martin--Siggia--Rose response fields. All the needed integrations over their arguments $x=\{t,{\bf x}\}$ are assumed here and in similar expressions below. The field theoretic formulation means that various correlation and response functions of the stochastic problem~(\ref{FPE})~-- (\ref{noise}) are represented by functional averages with weight $\exp {\cal S}(\Phi)$. 
The constant $D_0$ can be removed by rescaling the fields $h$, $h'$ and the  parameters. Thus, with no loss of generality, we set $D_0=1$.

In the following, an important role will be played by the Galilean-type transformation
\begin{equation}
t \to t'=t, \quad {\bf x}\to {\bf x}'=  {\bf x} + \lambda_0 {\bf u}t
\label{Galileo}
\end{equation}
with a constant arbitrary parameter ${\bf u}$. 
The equation~(\ref{equation1}), (\ref{KPZ}) is covariant with respect to this transformation if the field $h$ transforms as
\begin{equation}
h(t, {\bf x}) \to h(t, {\bf x} +  \lambda_0{\bf u}t) 
+ ({\bf u}{\bf x}) t\, +  {u^2\,t}/{2},
\label{galileo}
\end{equation}
which for the vector field $v_i=\partial_i h$ reduces to a more familiar form
\begin{equation}
{\bf v} (t, {\bf x}) \to {\bf v}(t, {\bf x} + \lambda_0{\bf u} t)
+ {\bf u},
\label{galileo1}
\end{equation}
that for $\lambda_0=-1$ coincides with the ordinary Galilean transformation.

The action functional~(\ref{Actionh}) is invariant with respect to this transformation if the additional rule for the field $h'$ is assumed:
\begin{equation}
h'(t, {\bf x}) \to h'(t, {\bf x}+  \lambda_0{\bf u}t). 
\label{galileoa}
\end{equation}

In its turn, the functional~(\ref{ActionT}) is invariant with respect to the transformation~(\ref{Galileo}), (\ref{galileo1}) with the replacement $\lambda_0 \to \alpha_0$ and 
\begin{equation}
\theta(t, {\bf x}) \to \theta(t, {\bf x} + \alpha_0 {\bf u}\,t),
\label{galileo2}
\end{equation}
\begin{equation}
\theta'(t, {\bf x}) \to \theta' (t, {\bf x} + \alpha_0 {\bf u}\,t).
\label{galileo9}
\end{equation}

The vector formulation~(\ref{Burgers}), (\ref{Burgers2}) allows one to interpret ~(\ref{Kolya}), (\ref{Burgers3}) as the advection-diffusion equation  
for a scalar density field passively advected by a strongly compressible stirred fluid.
Then it is natural to put $\lambda_0=\alpha_0$ from the very beginning. 
Indeed, for this special choice of the parameters 
the derivatives~(\ref{Burgers2})
and~(\ref{Burgers3}) become different only by a covariant contribution,
$\nabla_t = \nabla_t' + \lambda_0 (\partial_k v_k)$ and
the total action functional~(\ref{Action}) for the full model becomes invariant under the unified transformation~(\ref{Galileo})~-- (\ref{galileo9}).

For the application to random walks on a rough surface, the choice $\lambda_0=\alpha_0$ is by no means justified by the physical origin of these parameters: the former one
is related to the dynamics of the surface while the latter one involves the walking  particle's mass and the gravitational acceleration. 
However, as we will see, in a certain hypothetical strong-coupling IR regime, the Galilean symmetry emerges asymptotically in the Green's functions 
of the full model for arbitrary values of the bare parameters $\lambda_0$ and $\alpha_0$.

\section{Canonical dimensions, ultraviolet divergences, renormalization and one-loop results \label{QFT}}

Ultraviolet (UV) divergences can be analysed by considering canonical dimensions, see, e.g.~\cite{Vasiliev}, sections~1.15 and~1.16. As the model~(\ref{Action})~-- (\ref{Actionh}) involves time, i.e., it is dynamic rather than static, there are two independent scales: a time scale $[T]$ and a spatial scale $[L]$; see~\cite{Vasiliev}, sections~1.17 and~5.14. Canonical dimension of a quantity $F$ (field or parameter) is then determined by the frequency dimension $d_F^\omega$ and the momentum dimension $d_F^k$: 
\begin{equation} 
\left[F\right] \sim \left[T\right]^{-d_F^\omega} \left[L\right]^{-d_F^k}.
\nonumber
\end{equation}
A number of normalization conditions is assumed:
\begin{equation}
	d_{\bf k}^k = -d_{\bf x}^k = 1, \quad d_{\bf k}^\omega = d_{\bf x}^\omega = 0, \quad d_\omega^k = d_t^k = 0, \quad d_{\omega}^{\omega}=-d_t^{\omega}=1.
	\nonumber
\end{equation}
Taking into account that all terms in the action functional are dimensionless with respect to each dimension, we are able to calculate all canonical dimensions.

The total canonical dimension  is defined as $d_{F} = d_{F}^{k} + 2d_{F}^{\omega}$, the factor $2$ is due to the dimensional relation $\partial_t \propto {\bf \partial}^2$ in the free theory. While only momentum dimension is needed to analyse UV divergences in static models, $d_{F}$ must be introduced here to account for the frequency dimension contribution; see section~5.14 in~\cite{Vasiliev}.

Canonical dimensions of the fields and the parameters in our model are given in Table~\ref{t1}. Renormalized parameters (without subscript  ``0'') and the reference scale $\mu$ (renormalization mass) are also included; they will be introduced later. 

Note that for the fields $\theta'$, $\theta$, it is impossible to unambiguously define separate canonical dimensions so instead we list those for the product $\theta'\theta$. This is due to the fact that all terms of the action  functional~(\ref{Action}) contain equal number of the fields $\theta'$, $\theta$. More formally, this follows from the invariance of~(\ref{Action}) under the dilatation 
$\theta'\to\lambda\theta'$, $\theta\to\lambda^{-1}\theta$.

\begin{center}
\begin{table}[h!]
\centering
\caption{Canonical dimensions in the model~(\ref{Action})~-- (\ref{Actionh}).}
\label{canonical dimensions}
% \begin{ruledtabular}
\begin{tabular}{|c||c|c|c|c|c|c|c|c|}
 \hline
$F$&  $\theta'\theta$ & $h'$ & $h$ & $\sigma_0$, $\sigma$,  $\kappa_0$, $\kappa$ & 
$\alpha_{0}$, $\lambda_{0}$ & $g_{0}$, $w_{0}$ & $u_0$, $u$,
$g$, $w$  &{$\mu$,$m$}\\
\hline\hline
$d^{k}_F$ & $d$ & $d/2$ & $d/2$ & $-2$ & $-2-d/2$ & $1-d/2$ & $0$ & $1$\\
 \hline
$d^{\omega}_F$ & $0$ &$1/2$ &$-1/2$& $1$ & $3/2$& $0$ & $0$ & $0$\\
 \hline
$d_F$ & $d$ & $d/2+1$ & $d/2-1$ & $0$ & $1-d/2$ & $1-d/2$ & $0$ & $1$\\ 
 \hline
\end{tabular}
% \end{ruledtabular}
\label{t1}
\end{table}
\end{center}

For the following, it is convenient to pass to the new parameters $g_0=\lambda_0 \sigma_0^{-3/2}$ and
$w_0=\alpha_0 \sigma_0^{-3/2}$ that play the roles of coupling constants (expansion parameters in ordinary perturbation theory). 
The dimensionless ratio $u_0=\kappa_0/\sigma_0$ is not an expansion parameter, but its renormalized counterpart $u$ enters the renormalization constants and RG functions 
along with $g$ and $w$, therefore, it should be treated on equal footing with the couplings $g_0$ and $w_0$.

As can be seen from Table~\ref{t1}, the model becomes logarithmic (all the couplings appear dimensionless) at $d=2$; thus, the UV divergences in the Green's functions acquire the forms of poles in the parameter $\epsilon=2-d$ that measures ``deviation from logarithmicity.''

The total canonical dimension of a certain 1-irreducible Green's function is given by 
\begin{equation}
d_{\Gamma} = (d+2) - \sum_{\Phi} d_{\Phi} N_{\Phi},
 \label{Index}
\end{equation}
where $\Phi$ stands for the fields entering the Green's function,
$N_{\Phi}$ are the number of these fields, and $d_{\Phi}$ are their total canonical dimensions.
In our model, $N_{\theta}$ equals $N_{\theta'}$ for all nonvanishing Green's function (see the remark above Table~\ref{t1}). Thus, expression~(\ref{Index}) can be written as
\begin{equation}
d_{\Gamma} = (d+2) - d_{\theta'\theta} N_{\theta'\theta} - d_{h'} N_{h'} - d_{h} N_{h},
 \label{index}
\end{equation}
where $N_{\theta'\theta}$ is the number of the products $\theta'\theta$ entering the function.

Formal index of divergence $\delta_{\Gamma}$ of Green's function ${\Gamma}$ is the total dimension of that function taken in the logarithmic theory where $\epsilon=0$ or, equivalently, $d=2$. If $\delta_{\Gamma}$ is a nonnegative integer then there are superficial UV divergences in the Green's function $\Gamma$ and counterterms must be introduced to ``weed them out''.

When analyzing the divergences in the model~(\ref{Action})~-- (\ref{Actionh}), the following additional considerations should be taken into account; see, e.g.~\cite{Vasiliev}, section~5.15, \cite{UFN}, and~\cite{Red}, section~1.4. 

(i) All the 1-irreducible functions without the response fields contain closed circuits of retarded propagators
$\langle\Phi\Phi'\rangle_0$ and vanish. Thus, it is sufficient to consider the functions with $N_{\theta'}+N_{h'} >0$.

(  ii) One of the spatial derivatives $\partial_i$ in vertex $\theta'\partial_i (\theta \partial_i h)$ in~(\ref{Action})
can be transferred onto the field $\theta'$ via integration by parts: $\theta'\partial_i (\theta \partial_i h) \simeq -(\partial_i \theta')(\partial_i h)\theta$. Moreover, in both vertices $\theta'\partial_i (\theta \partial_i h)$ and $h' (\partial h)^2 h$, the field $h$ is under spatial derivative. Thus, in any 1-irreducible diagram, each external field $\theta'$ or $h'$ produces an  external momentum decreasing the real index of divergence by the corresponding number: $\delta'=\delta-N_{\theta'}-N_{h}$. It also means that these fields can only appear in counterterms under spatial gradients. This is why the counterterms $h'\partial_t h$,  $\theta'\partial_t \theta$
and $(\theta'\theta)^2$ are forbidden despite being allowed by the formal index of divergence.

All of these considerations combined result in the following expressions for the formal index of divergence and for the real one:
\begin{equation}
\delta= 4-2N_{h'}-2N_{\theta'\theta},  \quad
\delta'= 4 -2N_{h'}-N_{h}- 3 N_{\theta'\theta}.
\label{deltas}    
\end{equation}

Here it is also important to note that the ``submodel'' ${\cal S}_h$ in~(\ref{Action}) is multiplicatively renormalizable in itself, a well-known fact; see~\cite{Wiese2} and the references therein. Its renormalized action has the form
\begin{eqnarray}
{\cal S}_{hR}(h',h) = \frac{1}{2} Z_1 h'h'+ h'\left[-\partial_t h + Z_2 \sigma \partial^{2}h + g\sigma^{3/2}\mu^{\varepsilon/2} (\partial h)^2/2 \right],
\label{ActionhR}
\end{eqnarray}
where $Z_1$ and $Z_2$ are the renormalization constants, and all the fields and parameters are substituted with their renormalized counterparts.

The action~(\ref{Actionh}) is invariant with respect to the transformation~(\ref{Galileo}), (\ref{galileo}), (\ref{galileoa}). This excludes the counterterm 
$h'\,(\partial h)^2$ allowed by the real index of divergence.
What is more, specific features of the KPZ model allow one to show that the leading-order expressions for the renormalization constants in the minimal subtraction (MS) scheme
\begin{eqnarray}
Z_1^{-1} = 1+ \frac{1}{16\pi}\, \frac{g^2}{\epsilon}, \quad Z_2 = 1
\label{102} 
\end{eqnarray}
are in fact exact in the following sense: they have no higher-order corrections in $g$, which is why we presented $Z_1^{-1}$ rather than $Z_1$; see~\cite{Wiese2}~-- \cite{Wiese} and references therein.

The full renormalized action has the form
\begin{eqnarray}
{\cal S}_{R}(\Phi) = \theta' \left[-\partial_t \theta + Z_3 u\sigma \partial^2\theta 
+  Z_4 w \sigma^{3/2} \mu^{\varepsilon/2}\partial(\theta \partial h) \right] + {\cal S}_{hR}(h',h)
\label{ActionR}
\end{eqnarray}
with ${\cal S}_{hR}(h',h)$ from~(\ref{ActionhR}) with the same renormalization constants $Z_{1,2}$.\footnote{From the original stochastic formulation~(\ref{FPE}), (\ref{equation1}), (\ref{noise}) 
it is clear that the field $\theta$ is passive in the sense that it does not affect dynamics of the field $h$.}

The fields and the parameters are related to their renormalized counterparts as follows:
\begin{equation}
    q_0=qZ_q \quad \mbox{for} \quad q=\{h,h',\theta,\theta',\sigma,\kappa\},
    \label{renP}
\end{equation}
\begin{equation}
    g_0=g Z_g \mu^{\varepsilon/2},
    \quad
    w_0 = w Z_w \mu^{\varepsilon/2}, \quad
    u_0 = u Z_u. 
    \label{renC}
\end{equation}
The renormalization constants for the fields and the parameters can be expressed in terms of the constants $Z_i$, $i=1, \dots, 4$
from the action functionals~(\ref{ActionhR}) and~(\ref{ActionR}): 
\begin{eqnarray}
         Z_h = Z_1^{-1/2}, \quad
        Z_{h'} = Z_1^{1/2}, \quad
         Z_{\theta' \theta} =1,  \label{100}  \\       
         Z_w = Z_4 \,Z_1^{1/2}\, Z_2^{-3/2}, \quad
         Z_u = Z_3 \,Z_2, \quad
         Z_g = Z_1^{1/2} \,Z_2^{-3/2}, \label{101}  \\  
         Z_{\kappa}= Z_3, \quad 
         Z_{\sigma} = Z_2,  
         \label{Zzz}
\end{eqnarray}
which, in their turn, can be calculated within the perturbation theory. The first-order calculation in the MS scheme gives:
\begin{eqnarray}
Z_3 = 1+ \frac{1}{8\pi}\, \frac{w^2}{\epsilon}\, \frac{(u-1)}{u(u+1)^2}+\dots, 
\quad
Z_4 = 1+ \frac{1}{8\pi}\, \frac{w}{\epsilon}\, \frac{(w-g)}{(u+1)^2}+ \dots\, . 
\label{Zs}
\end{eqnarray}
In contrast to~(\ref{102}), the constants~(\ref{Zs}) may have higher-order corrections in $g$ and $w$. However, for the case $w=g$, one has $Z_4=1$ identically to all orders of the perturbation expansion due to 
resulting reinstatement of the symmetry~(\ref{Galileo}), (\ref{galileo1}), (\ref{galileo2}), (\ref{galileo9}) for the full model~(\ref{Action}) that rules out the counterterm $\theta'\partial(\theta\partial h)$.

\section{RG equation, RG functions and fixed points of the RG equation \label{RGEq}}

For a multiplicatively renormalizable model the RG equations
can be derived in a standard fashion; see, e.g.~\cite{Vasiliev}. For the 
model~(\ref{ActionR}) one obtains:
\begin{equation}
\left\{ {\cal D}_{\mu} - \beta_{g} \partial_g - \beta_{w} \partial_{w} - \beta_u \partial_{u} - \gamma_{\sigma}  {\cal D}_{\sigma}
+ \gamma_{G} \right\}\, G(e;\dots)=0.
\label{RGE}
\end{equation}
Here $G(\cdot)$ is a certain renormalized Green's function of the model~(\ref{ActionR}) expressed in terms of the full set of renormalized variables $e=\left\{g,w,u,\sigma,\mu \right\}$. The ellipsis denotes other variables such as coordinates (or momenta) and times (or frequencies).
Here and below $\partial_x = \partial/\partial x$ and ${\cal D}_x = x \partial_x$ for any variable $x$.

The operation $\widetilde{{\cal D}}_{\mu}$
\begin{equation}
    \widetilde{{\cal D}}_{\mu} \equiv  \mu
   \frac{\partial}{\partial \mu}\Big|_{e_0}
\label{diffop}
\end{equation}
is taken at fixed bare parameters $e_0=\left\{g_0,w_0,u_0,\sigma_0\right\}$. 
The RG functions $\beta$ and $\gamma$ serve as coefficients in the RG equation~(\ref{RGE}) and are defined as follows:
\begin{equation}
    \gamma_a \equiv \widetilde{{\cal D}}_{\mu} \ln{Z_a}
\label{gammaf}
\end{equation}
for any quantity $a$ and 
\begin{equation}
    \beta_q \equiv \widetilde{{\cal D}}_{\mu} q = -q\,(d^k_{q_0} + \gamma_q)
\label{betaf}
\end{equation}
for any coupling constant $q$.

From the definition~(\ref{gammaf}) and the expressions for the renormalization constants~(\ref{102}) and~(\ref{Zs}), one obtains for anomalous dimensions $\gamma$:
\begin{eqnarray}
    \gamma_1 = \frac{g^2}{16\pi}, \quad 
    \gamma_2 = 0,  \quad 
    \gamma_3 = -\frac{w^2}{8\pi} \frac{(u-1)}{u(u+1)^2}+\dots, \label{108} \\
    \gamma_4 = -\frac{w(w-g)}{8\pi} \frac{1}{(u+1)^2}+\dots .
    \label{gamma-1}
\end{eqnarray}
The expressions for $\gamma_{1,2}$ are exact to all orders of the perturbation theory; see the remark below equation~(\ref{102}). The expressions for $\gamma_{3,4}$ are only one-loop approximations; however, the relation  $\gamma_{4}=0$ for $g=w$ is also exact; see the remark below equation~(\ref{Zs}).
Now using the relations~(\ref{Zzz}) one can calculate the anomalous dimensions of the coupling constants 
\begin{eqnarray}
\gamma_g = \frac{1}{2} \gamma_1 = \frac{g^2}{32\pi}, \quad \gamma_u = \gamma_3 = - \frac{w^2}{8\pi} \frac{(u-1)}{u(u+1)^2}+\dots, 
\label{110} \\
    \gamma_w = \gamma_4 + \frac{1}{2} \gamma_1 = \frac{g^2}{32\pi} - \frac{w(w-g)}{8\pi} \frac{1}{(u+1)^2}+\dots
    \label{gamma-2}
\end{eqnarray}
and for the other parameters:
\begin{equation}
    %\label{gammanu}
    \gamma_{\sigma} = \gamma_2 = 0, \quad \gamma_{\kappa}=\gamma_3=- \frac{w^2}{8\pi} \frac{(u-1)}{u(u+1)^2}+\dots .
    \label{gamma-0}
\end{equation}
Thus, the $\beta$ functions~(\ref{betaf}) have the forms:
\begin{eqnarray}
    \beta_g = -g \left(d^k_{g_0} + \gamma_g \right) = 
    -g \left( \frac{\varepsilon}{2} + \frac{1}{2} \frac{g^2}{16\pi} \right), 
    \label{betag}
    \\
    \beta_w = -w \left(d^k_{w_0} + \gamma_w \right) = 
    -w \left( \frac{\varepsilon}{2} + \frac{1}{2}\, \frac{g^2}{16\pi} - \frac{w(w-g)}{8\pi}\, \frac{1}{(u+1)^2} \right) + \dots,
    \label{betaw}
    \\
    \beta_u = -u \left( d^k_{u_0} + \gamma_u \right) = 
    \frac{w^2}{8\pi}\, \frac{(u-1)}{(u+1)^2}+ \dots .
    \label{betau}
\end{eqnarray} 
Again, the expression~(\ref{betag}) is exact in perturbation theory.

Fixed points of the RG equation define regimes of scaling behaviour, and their coordinates $\{g_*,w_*,u_*\}$ are given by the solutions of the equation system 
\begin{equation}
\beta_g(g_*,w_*,u_*)=0,\quad \beta_w(g_*,w_*,u_*)=0,\quad \beta_u(g_*,w_*,u_*)=0.
\label{FPs}
\end{equation}
The character of a fixed point is determined by the stability matrix $\Omega_{ik}$ made out of the $\beta$-functions first derivatives:
\begin{equation}
    \Omega_{ik} \equiv \partial_i \beta_k (g_*,w_*,u_*).
\label{matrix}
\end{equation}
The point is IR attractive (or IR stable) if all the real parts of the eigenvalues
of the matrix $\Omega_{ik}$  are positive. 

There are four fixed points for the system~(\ref{betag})~--  (\ref{betau}) for which $u_*\neq 0$, $u_*\neq\infty$; however, due to the fact that $u_0$ is dimensionless ratio, these marginal cases must also be considered. To do that one can pass to another set of coupling constants, namely, $\{g_*, w_*, y_* = 1/u_*\}$ and $\{g_*, \widetilde{w}_* = w_* / u_*^{3/2}, u_*\}$. Each of these systems produces one new fixed point, so there are six in total:

\begin{enumerate}
    \item\label{ik1} The Gaussian fixed point with the eigenvalues 
    $\{ -{\varepsilon}/{2}, -{\varepsilon}/{2}, 0 \}$:
    \begin{equation}
        g_* = 0,
        \quad
        w_* = 0,
        \quad
        \mbox{any~$u_*$};
        \label{fp1}
    \end{equation}

    \item The fixed point that corresponds to the regime where the non-linearity in the KPZ equation~(\ref{KPZ}) is irrelevant, with the eigenvalues $\{ -{\varepsilon}/{2}, \varepsilon, {\varepsilon}/{2} \}$:
    \begin{equation}
        \label{fp2}
        g_* = 0,
        \quad
        w_*^2 = 16\,\pi \varepsilon,
        \quad
        u_* = 1;
    \end{equation}

    \item\label{ik3}
    The fixed point that corresponds to the regime where the interaction of the fields is irrelevant:
    \begin{equation}
        \label{fp3}
        g_*^2 = -16\pi \varepsilon,
        \quad
        w_* = 0,
        \quad
        \mbox{any~$u_*$};
    \end{equation}

    \item  The fixed point that corresponds to fully nontrivial regime, with the eigenvalues  $\{ \varepsilon, -{\varepsilon}/{2}, -{\varepsilon}/{2} \}$:
    \begin{equation}
        \label{fp4}
        g_*^2 = -16\,\pi \varepsilon, \quad w_* = g_*, \quad u_* = 1;
    \end{equation}

 \item\label{inf5} The fixed point from the system of charges $\{g_*, w_*, y_* = {1}/{u_*}\}$ that accounts for the limit  $y_*\to 0$:
    \begin{equation}
        \label{fp5}
        y_* = 0,
        \quad
        g_*^2 = -16\pi \varepsilon,
        \quad
        \mbox{any~$w_*$};
    \end{equation}

    \item\label{inf6}
    The fixed point from the system of charges $\{g_*, \widetilde{w}_* = w_* / u_*^{3/2}, u_*\}$ that accounts for the limit $u_* \to 0$:
    \begin{equation}
        \label{fp6}
        u_* = 0,
        \quad
        g_*^2 = -16\,\pi \varepsilon,
        \quad
        \mbox{any~$\widetilde{w}_*$}.
    \end{equation}
\end{enumerate}
The eigenvalues for the points~(\ref{ik3}), (\ref{inf5}), (\ref{inf6})
are given by $\{\varepsilon$, $0$, $0\}$. 
One of the vanishing eigenvalues is explained by the fact that these points are degenerate: they are actually lines of fixed points due to the arbitrariness of their coordinates~$u_*$, $w_*$ and~$\widetilde{w}_*$, respectively.
The second vanishing eigenvalue is due to the fact that 
one of the coordinates of these fixed points is a third-order root for the system of equations~(\ref{FPs}). 
In such cases, the role of the first derivative is conveyed to the third-order derivative (the first non-vanishing one) whose sign then determines stability of the fixed point along the corresponding direction; see, e.g. section~1.31 in~\cite{Vasiliev}.
The calculation shows that the points~(\ref{inf5}) and~(\ref{inf6}) are neither IR nor UV attractive for any value of~$\varepsilon$, and from the entire set of fixed points only points~(\ref{ik1}) and~(\ref{ik3}) are IR attractive for $\varepsilon<0$ and $\varepsilon>0$, respectively.

\section{Scaling behaviour and critical dimensions \label{Scaling}}

Critical dimension of a quantity $F$ (a field or a parameter) is given by the following expression (see, e.g. sections~6.7 in~\cite{Vasiliev}, \cite{UFN} and section~2.1 in~\cite{Red}):
 \begin{equation}
    \Delta_F = d_F^k + \Delta_{\omega} d^{\omega}_F + \gamma^*_F, 
    \qquad \Delta_{\omega} = 2 - \gamma^*_{\sigma}.
\label{critdim}
\end{equation}
Here $d_F^k$ and $d^{\omega}_F$ are canonical dimensions from Table 1, $\gamma^*_F \equiv \gamma_F (g_*,w_*,u_*)$, and the normalization condition is $\Delta_{k}=-\Delta_{x}=1$. 

For the Gaussian point~(\ref{ik1}), the critical dimensions 
\begin{equation}
    \label{critdim1}
    \Delta_h = {d}/{2} - 1, \quad \Delta_{h'} = {d}/{2} + 1, \quad \Delta_{\theta \theta'} = d, \quad \Delta_{\omega} = 2
\end{equation}
coincide with their canonical values and therefore are all exact, that is, they have no corrections of the order $\varepsilon^2$ and higher (we recall that $d=2-\varepsilon$).
For the point~(\ref{ik3}), the expressions
\begin{equation}
    \label{critdim3}
    \Delta_h = 0, \quad \Delta_{h'} = d, \quad \Delta_{\theta \theta'} = d, \quad \Delta_{\omega}=2
\end{equation}
are also exact. 
For the fields~$h$ and~$h'$ this is due to the relations between their anomalous dimensions and the anomalous dimension~$\gamma_1$ which is known exactly; see equations~(\ref{Zzz}), (\ref{gamma-1}). For the frequency $\omega$, this follows from the fact that $\gamma_{\sigma}$ vanishes identically; see equation~(\ref{gamma-0}) and explanation below~(\ref{Zs}). 
These two features are specific of the KPZ model~\cite{Wiese2}~-- \cite{Wiese}. 
Finally, since the product $\theta \theta'$ is not renormalized, $Z_{\theta \theta'}=1$ (see~(\ref{ActionR}) and~(\ref{Zzz})), the corresponding anomalous dimension $\gamma_{\theta \theta'}$ vanishes identically and the critical dimension of the product $\Delta_{\theta \theta'}$ coincides with its canonical value for both fixed points~(\ref{ik1}) and~(\ref{ik3}).

A remark is in order here. The full stochastic problem~(\ref{equation1})~-- (\ref{noise}), (\ref{Kolya}), (\ref{Burgers3}) involves two parameters $\sigma_0$ and $\kappa_0$ with the same canonical dimensions and 
the dimensionless ratio $u_0$. This produces an ambiguity in the choice of the coupling constants like $g_0$ and $w_0$.   
In particular, the frequency critical dimension $\Delta_{\omega}$ can be defined as either $\Delta_{\omega}=2-\gamma^*_{\sigma}$ as we done in~(\ref{critdim}) or $\Delta_{\omega}=2-\gamma^*_{\kappa}$. For the fixed points~(\ref{ik1}) and~(\ref{ik3}) with $w_*=0$, this means that the fields $h$, $h'$ and $\theta$, $\theta'$ 
decouple (in the asymptotic IR range)
and the full problem reduces to the two independent separate sub-problems~(\ref{equation1})~-- (\ref{noise}) and~(\ref{Kolya}), (\ref{Burgers3}). In cases like this, the ambiguity is resolved by each sub-problem getting its own frequency critical dimension; the phenomenon referred to as weak scaling~\cite{Folk}.

This issue, however, is irrelevant here as both definitions lead to the same result. Since all the nontrivial diagrams entering the Green's functions $\langle \theta\theta \dots \theta'\theta' \rangle_{\mbox{1-ir}}$ contain vertices $\theta'\theta h$, they effectively vanish at~(\ref{ik3}) where $w_*=0$. This means that only the loopless diagrams survive and thus, $\gamma^*_{\kappa}$ vanishes exactly at~(\ref{ik3})
along with $\gamma^*_{\sigma}$.

\section{Non-perturbative RG treatment of the KPZ model and its implications \label{KPZ-NP}}

As we could see from the preceding sections, the full model~(\ref{Action})~-- (\ref{Actionh}) does not have nontrivial IR fixed points in the physical range of parameters. ``Unconventional'' or ``unphysical'' points, including complex, imaginary or unstable ones, admit various interesting interpretations; see, e.g.~\cite{Yakhot}~-- \cite{AntKo}. However, it would be tempting to identify more conventional ``admissible'' fixed points that would be IR attractive and belong to the physical range. Of course, the key origin of the problem lies in the inherent peculiarity of the pure KPZ model, as explained below.

Stochastic KPZ equation~(\ref{equation1})~-- (\ref{noise}) got the RG treatment in the papers it was introduced in~\cite{KPZ}~-- \cite{FNS1} through Wilson's recursion relations; later, it was investigated with the more advanced field theoretic RG that opened the door for higher-order calculations and more nuanced technical analysis; see~\cite{Wiese2} for the references and discussion. 

The original stochastic KPZ problem~(\ref{equation1})~-- (\ref{noise}) was reformulated as a multiplicatively renormalizable field theoretic model~(\ref{Actionh}) with logarithmic value $d=2$, the RG expansion parameter $\varepsilon=2-d$ and the corresponding renormalized action~(\ref{ActionhR}) with the exactly known renormalization constants $Z_{1,2}$ 
given in~(\ref{Zs}).

For the readers's convenience, here we pass to the more traditional notation
$\Delta_{h}=-\chi$ and $\Delta_{\omega}=z$ and to the new coupling constant
$q=g^2/16\pi$. Then the function~(\ref{betag}) turns to
\begin{equation}
\beta_q = -q\, (\varepsilon+q).
\label{betaKPZe}    
\end{equation}
We recall that this expression in the MS scheme is exact within the perturbative expansion, in the sense that it has no higher-order corrections in $q$.
We also apologize to the reader for recalling that possible asymptotic scaling regimes are governed by the fixed points $q^*$, solutions of the equation $\beta(q^*) =0$. 
The point is IR attractive for $\beta'(q^*)>0$ and IR repulsive (UV attractive) for $\beta'(q^*)<0$.

From~(\ref{betaKPZe}) it follows that the trivial (free or Gaussian) fixed point $q^*=0$ is IR repulsive for $d<2$ and attractive for $d>2$. For this point, the critical exponents are found exactly, $\chi=(2-d)/2$, $z=2$, and correspond to a smooth surface.

The nontrivial fixed point $q^*=-\varepsilon$ is IR attractive for $d<2$ but lies in the unphysical region $q^*<0$. For $d>2$, this point transfers to the physical region $q^*>0$ and becomes IR repulsive. It is usually interpreted as the boundary between the interval $0<q<q^*$, where the IR behaviour is governed by the Gaussian point $q^*=0$, and the region $q>q^*$ with no fixed point.
Thus, the standard perturbative RG fails to give any prediction for the IR asymptotic behaviour for any $q>0$, $\varepsilon>0$ and for $q>-\varepsilon>0$.

However, various non-perturbative considerations (numerical simulations, mode-coupling theories, mapping onto directed polymers, functional RG, fractal description) imply
existence of a strong-coupling scaling regime (rough phase) for all $1\le d < d_c$, where $d_c$ is the upper critical dimension; see, e.g.~\cite{Kinzel}~-- \cite{Canet4} and references therein.\footnote{In early works~\cite{Wiese2}, \cite{Kinzel}, it was argued that $d_c\le 4$, while more recent studies suggest that $d_c=\infty$; \mbox{see~\cite{Gomes}, \cite{Katzav}, \cite{Upperinf}.}}

Within the RG framework, it is natural to associate this  strong-coupling regime with a certain non-perturbative IR attractive fixed point, not ``visible'' in the perturbative expression~(\ref{betaKPZe}). Existence of this fixed point is supported by the functional RG~\cite{Canet}~-- \cite{Canet4}. This point governs the IR behaviour for all $q>0$ if $\varepsilon>0$ and for $q>-\varepsilon>0$ if $\varepsilon<0$. For $\varepsilon<0$ and 
 $0< q <-\varepsilon>0$ the IR behaviour is determined by the Gaussian point $q=0$ (smooth phase). Thus, the nontrivial perturbative fixed point $q_*=-\varepsilon$ plays a role of the boundary between the two phases; see, e.g. the discussion in~\cite{Kinzel} and references therein\footnote{Interestingly, this pattern of fixed points of the KPZ equation is simulated by a slight change of~(\ref{betaKPZe}) to $\beta(q) = -q\, (\varepsilon+q-a q^2)$ with $a>0$~\cite{Tumakova}. This form even implies the finite value of upper critical dimension $d_c = 2-1/4a\varepsilon >2$. 
At $d=d_c$,  corresponding strong-coupling fixed point and perturbative IR repulsive fixed point fuse together, disappearing for $d>d_c$. This makes the Gaussian point $q^*=0$ the one that decides the IR behaviour for all $q>0$. The $\beta$ function of this form was obtained for $d=2$ in~\cite{Teodor} where a non-minimal renormalization scheme was employed.}.

For general $d$, the Galilean symmetry~(\ref{Galileo})~-- (\ref{galileoa})
excludes the counterterm $h'\nabla_t h$; this leads to the exact expression $\chi+z=2$ for any nontrivial fixed point; see~\cite{FNS1} and sections~\ref{QFT} and~\ref{Scaling}.
For $d=1$, the fluctuation-dissipation relation makes it possible to find the exact values $\chi=1/2$, $z=3/2$~\cite{FNS1}.

A number of works attempted at establishing exact results for the case $d>1$. In particular, in the paper~\cite{Lassig}, the operator-product expansion coupled with certain non-perturbative considerations produced an infinite sequence of exact values for $\chi$ (and thus, for $z$ as well). Two couples among them were identified as the exponents for $d=2$ and $d=3$. Fractal framework was employed in~\cite{Gomes} where by establishing fractal dimension of a growing (rough) surface it was possible to find exact values of the exponents. These results are in reasonable agreement with numerical simulations; see, e.g.~\cite{Kinzel}, \cite{Tang}~-- \cite{Ala1}, section~4 in~\cite{Gomes} and references therein.

In the following, we explore implications of that non-perturbative fixed point existence on the behaviour of the density field $\theta$
in the full-fledged model~(\ref{Action})~-- (\ref{Actionh}). The only assumption we make is that the ``genuine'' anomalous dimension $\gamma_1=\gamma_g$ involves a certain non-perturbative contribution such that the equation $\beta_g=\beta_q=0$ has the desired fixed point as a solution. At the same time, all the relations between the $\beta$ and $\gamma$ functions as well as the explicit one-loop expressions for the anomalous dimensions $\gamma_{3,4}$ are taken from our standard perturbative RG analysis; see~(\ref{gamma-1})~-- (\ref{betau}).

Then it is clear that the equation $\beta_w=0$ (with the assumption that $w \ne 0$) leads to $\gamma_4 =0$; see equations~(\ref{gamma-2}), (\ref{betag}) and~(\ref{betag}).
The dimension $\gamma_4$ vanishes identically for $w=g$ due to the Galilean symmetry
that takes place for this special choice; see the remark under equation~(\ref{Zs}). Thus, we arrive at the exact solution $w_*= g_*$ with $g_*^2=w_*^2>0$. Then the equation $\beta_u=0$ provides $u_*=1$ in the one-loop approximation~(\ref{betau}).

The resulting ``hybrid'' fixed point appears IR attractive. Indeed, by assumption, $\Omega_{gg} =\partial_g \beta_g >0$. Then, the derivatives $\partial_u \beta_g = \partial_w \beta_g =0$ vanish due to the passivity of $\theta$ for any possible expression for $\beta_g$, while $\partial_w \beta_u  \sim (u-1) =0$ vanishes for $u_*=1$. Thus, the matrix~(\ref{matrix}) is block triangular and its eigenvalues coincide with the diagonal elements; from the expressions~(\ref{betaw}), (\ref{betau}) one obtains $\partial_w \beta_w = \partial_u \beta_u = w^2_* /(32\pi)>0$.

Thus, unlike the case of the fully nontrivial perturbative fixed point~(\ref{fp4}),
this hybrid point is IR attractive, lies in the physical range of parameters and, therefore, governs the IR asymptotic behaviour of the Green's functions
of our model.

The critical dimensions $\Delta_{\omega}=z$, $\Delta_{h}=-\chi$ 
for this hybrid fixed point coincide with the ones of the pure KPZ model and can be inferred from existing studies~\cite{Gomes}~-- \cite{Ala1},\footnote{It is crucial here that $u_*=1$ is finite and the weak scaling~\cite{Folk} does not take place.}
while the exact value $\Delta_{\theta\theta'}=d$ results from the present RG analysis.

When discussing the non-perturbative regime, we confined ourselves with the exact solution $w_*=g_*$ of equation $\gamma_4=0$; 
as a result, there is a dynamic emergence of the symmetry~(\ref{Galileo}), (\ref{galileo1}), (\ref{galileo2}), (\ref{galileo9}) in that regime. To avoid possible misunderstanding, it should be stressed that the bare couplings
$g_0$ and $w_0$, as well as their renormalized counterparts $g$ and $w$, are kept fixed. It is the corresponding {\it invariant or running couplings} (special solutions of the RG equation; see, e.g. sections~1.29~-- 1.32 in~\cite{Vasiliev}), that asymptotically approach the fixed point with equal coordinates $w_*=g_*$.
In the one-loop approximation, this regime is also the one where non-linearity of the KPZ equation and the interaction between the fields $h$, $\theta$ and $\theta'$ are both relevant while the renormalized ratio $u_*$ has a finite non-zero value.

\section{Spreading law for particles' cloud} \label{Spread}

As a specific illustration of the application of the obtained results, let us consider the root-mean-square displacement $R(t)$ of the ``walker.''
 For a particle that starts at the moment $t=0$ from the origin ${\bf x}=0$, it is given by the following expression; see, e.g.~\cite{Vasiliev}, page~711:
\begin{equation}
 \label{cloud}
 R^2(t) = \int\, d{\bf x}\, x^2 \langle \theta(t,{\bf x})
 \,\theta'(0,{\bf 0}) \rangle,
\end{equation}
where $t>0$~is the  ``observation time'' and~${\bf x}$~is the corresponding position.

The IR asymptotic scaling law for the linear response function 
\begin{equation}
 \label{response}
 G (t,{\bf x}) = \langle \theta(t,{\bf x}) \, \theta'(0,{\bf 0})
 \rangle \simeq r^{-\Delta_{\theta\theta'}}\,
 F(tr^{-\Delta_{\omega}}), \quad r\equiv |{\bf x}|
\end{equation}
results in the following large-$t$ behaviour for $R^2$:
\begin{equation}
 \label{cloud9}
 R^2(t) \propto t^{(d+2-\Delta_{\theta\theta'})/\Delta_{\omega}}.
 \end{equation}
For all the fixed points discussed above, 
the exact relation $\Delta_{\theta'\theta}=d$ holds, so that~(\ref{cloud9}) becomes: 
\begin{equation}
 \label{cloud90}
 R^2(t) \propto t^{2/\Delta_{\omega}}.
 \end{equation}
For the perturbative fixed points~(\ref{ik1}) and~(\ref{ik3}) one has 
$\Delta_{\omega}=2$, and the spreading law~(\ref{cloud90})
takes on the form
\begin{equation}
 \label{cloud10}
 R^2(t) \propto t.
\end{equation}
This result coincides with the case of ordinary random walks. The reason, of course, is in very simple exact expressions~(\ref{critdim1}), (\ref{critdim3}) for critical dimensions.
In this respect, our model differs from the problem studied in~\cite{Dima}, where the surface was modelled by a generalized Gaussian Edwards-Wilkinson ensemble: while the critical dimensions $\Delta_{\omega}$ and $\Delta_{\theta'\theta}$ in the spreading law~(\ref{cloud9}) were also found exactly there, the former differed for different fixed points. However, in the present model, a nontrivial spreading law appears for the regime corresponding to the non-perturbative strong-coupling fixed point.

In that regime, $\Delta_{\theta'\theta}$ is again equal to $d$ and~$\Delta_{\omega}=z$. Thus, the spreading law is given by equation~(\ref{cloud90}). For $d=1$, one obtains $R^2(t) \propto t^{4/3}$
due to the fluctuation-dissipation relation.
For $d=2$, a range of possible values was reported: $z=8/5=1.6$, see~\cite{Lassig};
$z=(1+\sqrt{5})/2 \simeq 1.618$ (the golden mean), see~\cite{Gomes}; $z \simeq 1.612$, see~\cite{Gomes}~-- \cite{Ala1}. For $d=3$, the values $z=12/7 \simeq 1.714$, see~\cite{Lassig};
$z=\sqrt{3} \simeq 1.732$, see~\cite{Gomes}; $z \simeq 1.69$, see~\cite{Lassig}~-- \cite{Ala}, were proposed.
The numerical simulation of~\cite{Ala} suggest that $z\to2$
for $d\to\infty$, so that the spreading law~(\ref{cloud90})
approaches that for ordinary random walk.

At a first glance, it seems strange that the random walk on a rugged terrain proceeds faster than that on a smooth surface. In fact, the reason is that the surface is fluctuating in time. This effect can be compared with the diffusion in a strongly turbulent medium, where, according to the ``four-thirds'' Richardson law, $R^2(t) \propto t^{3}$; see, e.g.~\cite{McComb}.

\section{Conclusion \label{Conc}}

For the studied model of random walks~(\ref{FPE}),~(\ref{gravi}) on a growing surface described by the KPZ equation~(\ref{equation1})~-- (\ref{noise}), the action functional of quantum field theory~(\ref{Action})~-- (\ref{Actionh}) was constructed. Its UV divergences were analyzed by calculating canonical dimensions of the fields and~the parameters of the model (Table \ref{canonical dimensions}) allowing to establish that the action is multiplicatively renormalizable~(\ref{ActionhR}), (\ref{ActionR}). Renormalization constants~(\ref{Zs}) were found in one-loop approximation or exactly, see Appendix. 

It was established that RG equation~(\ref{RGE}) has six fixed points (or curves)~(\ref{fp1})~-- (\ref{fp6}); among them, only two can be IR-attractive and define types of asymptotic behaviour (universality classes): the Gaussian (free) point~(i) at~$\varepsilon<0$ and the point (iii) at $\varepsilon>0$. The latter point corresponds to the regime where only non-linearity of the KPZ equation is relevant; as a result, this point lies outside of the physical region of parameters which is a feature typical for the RG analysis of the KPZ equation. For these points, the critical dimensions of the fields and the parameters of the system~(\ref{critdim1}) and~(\ref{critdim3}) are calculated exactly.

Additionally, implications of the existence of non-perturbative (strong-coupling) fixed point for the behaviour of the density field $\theta$
in the full model (\ref{Action})~-- (\ref{Actionh}) were explored, see section~\ref{KPZ-NP}. It was shown that the fully nontrivial regime where both non-linearity of the KPZ equation and its interaction with the random walkers are relevant would be IR attractive for positive $\varepsilon$ with critical dimensions $\Delta_{\omega}=z$, $\Delta_{h}=-\chi$ and $\Delta_{\theta\theta'}=d$. In this regime, full Galilean symmetry~(\ref{Galileo}), (\ref{galileo1}), (\ref{galileo2}), (\ref{galileo9}) emerges asymptotically.

As an example of the application of the results, we considered the mean-square displacement of a walking particle~(\ref{cloud}) (in another interpretation, the radius of the particle cloud). For the perturbative fixed points, the law~(\ref{cloud10}) coincides with the case of ordinary random walks. The law for the non-perturbative regime
can be obtained by taking value of critical exponent $z$ for the KPZ equation from the literature;
it turns out that the random walk is faster when on a rough than when on a smooth one which is similar to the ``four-thirds'' Richardson law for the diffusion in a strongly turbulent medium.

To further study how asymptotic behaviour of random walks depend on the statistics of $h$, it would be interesting to consider conservative dynamics. From equilibrium critical dynamics, it is known that critical regimes strongly depend on whether the field is a density of some conserved quantity, i.e., whether the equation describing the system evolution has the form of a continuity equation. There are several versions of ``conserved'' KPZ equation~\cite{Modify10}~-- \cite{CKPZSQN} that can serve as models for the rough surface. This work is in progress.

\section*{Acknowledgments}

The work of P.~I. Kakin was supported by the Foundation for the Advancement of Theoretical Physics and Mathematics ``BASIS'' (project~22-1-3-33-1).
The work of N.~M.~Gulitskiy and A.~S.~Romanchuk was performed at the Saint Petersburg Leonhard Euler International Mathematical Institute and supported by the Ministry of Science and Higher Education of the Russian Federation (agreement~075–15–2022–287).

\section*{Appendix: Diagrammatic technique and one-loop calculation}

The propagators for the model~(\ref{Action})~-- (\ref{Actionh}) in the frequency-momentum representation have the forms:
\begin{equation} 
\begin{aligned}
\langle hh \rangle_0 &= \frac{1}{\omega^2+\sigma_0^2 k^4}, &
\langle hh' \rangle_0 &=
\langle h'h \rangle_0^{*} =
\frac{1}{-i\omega+\sigma_0 k^2},
&
\langle h'h'\rangle_0 &= 0,
\\
\langle \theta \theta \rangle_0 &= 0, &
\langle \theta \theta' \rangle_0 &= \langle \theta' \theta \rangle_0^{*}
=  
\frac{1}{-i\omega+\sigma_0 u_0 k^2}, &
\langle \theta' \theta' \rangle_0 &= 0. 
\label{propagators}
\end{aligned}
\end{equation}

In the Feynman diagrammatic technique for our model, the following lines are associated with the non-zero propagators in~(\ref{propagators}):
\begin{equation}
    \langle \theta \theta' \rangle_0 = \includegraphics[height=2ex, valign=c]{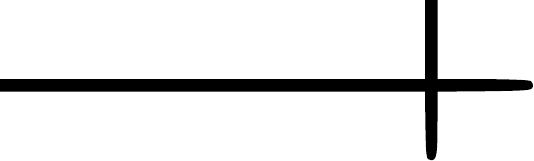}, \quad
        \langle hh' \rangle_0 = \includegraphics[height=2ex, valign=c]{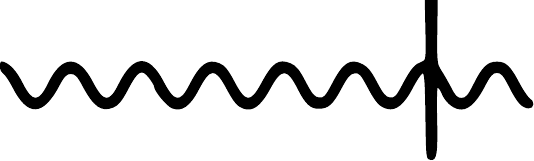}, \quad
    \langle hh \rangle_0 = \includegraphics[height=0.6ex, valign=c]{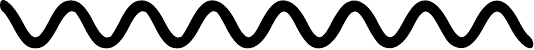}.
\label{lines}
\end{equation}
Interaction terms of the action~(\ref{Action})~-- (\ref{Actionh}) correspond to the vertices:
\begin{equation}
    \theta \theta' h = \includegraphics[height=8ex, valign=m]{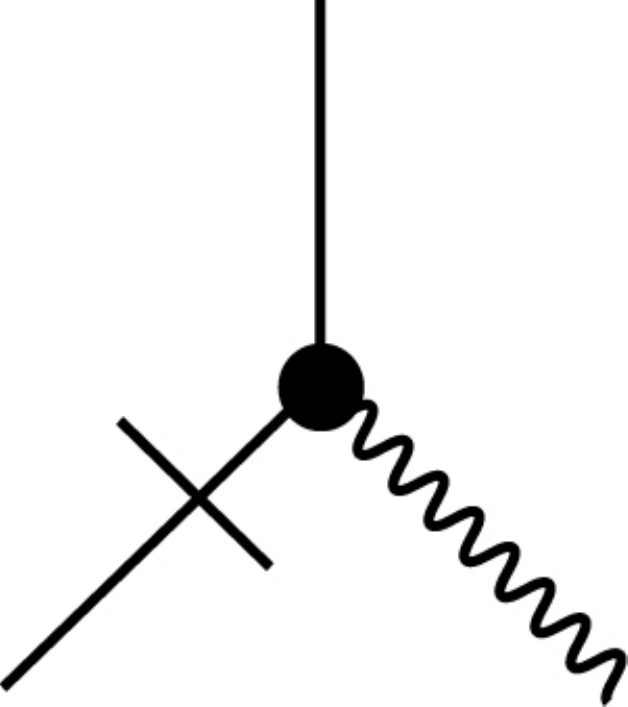},
    \qquad
    h'hh = \includegraphics[height=8ex, valign=m]{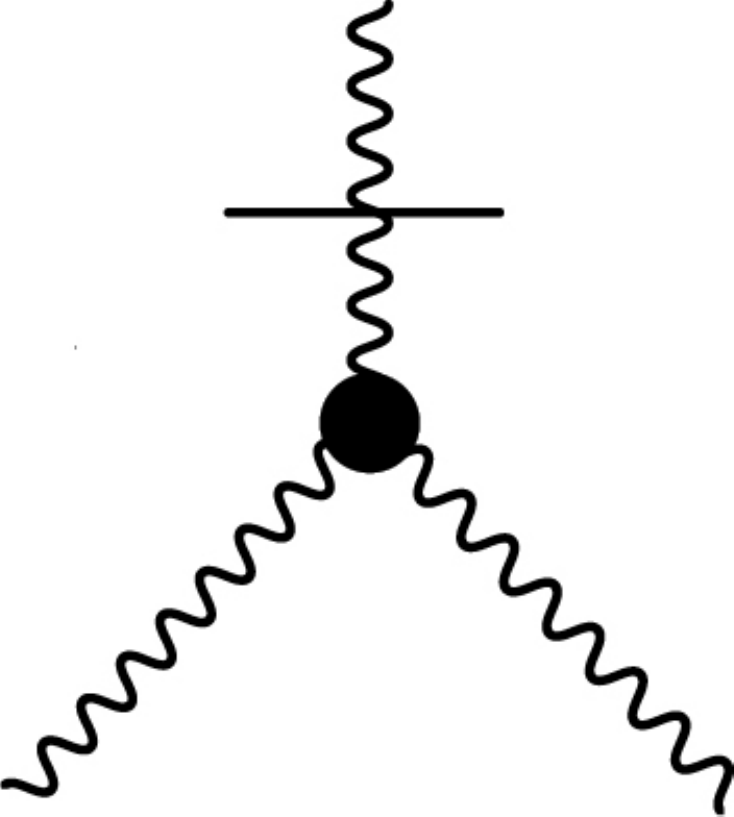}.
\label{vertices}
\end{equation}

The renormalization constants $Z_1$~-- $Z_4$ can be calculated from the four 
1-irreducible functions: $\langle h'h' \rangle_{\text{1-ir}}, \langle h'h \rangle_{\text{1-ir}}, \langle \theta \theta' \rangle_{\text{1-ir}}$ and~$\langle \theta \theta' h \rangle_{\text{1-ir}}$. In the one-loop approximation, they correspond to a total of six diagrams (${\bf q}$ and ${\bf p}$ being external momenta):
\begin{align}
 \langle h'h' \rangle_{\text{1-ir}} &= Z_1 +
    \frac{1}{2}\,
    \includegraphics[height=7ex, valign=c]{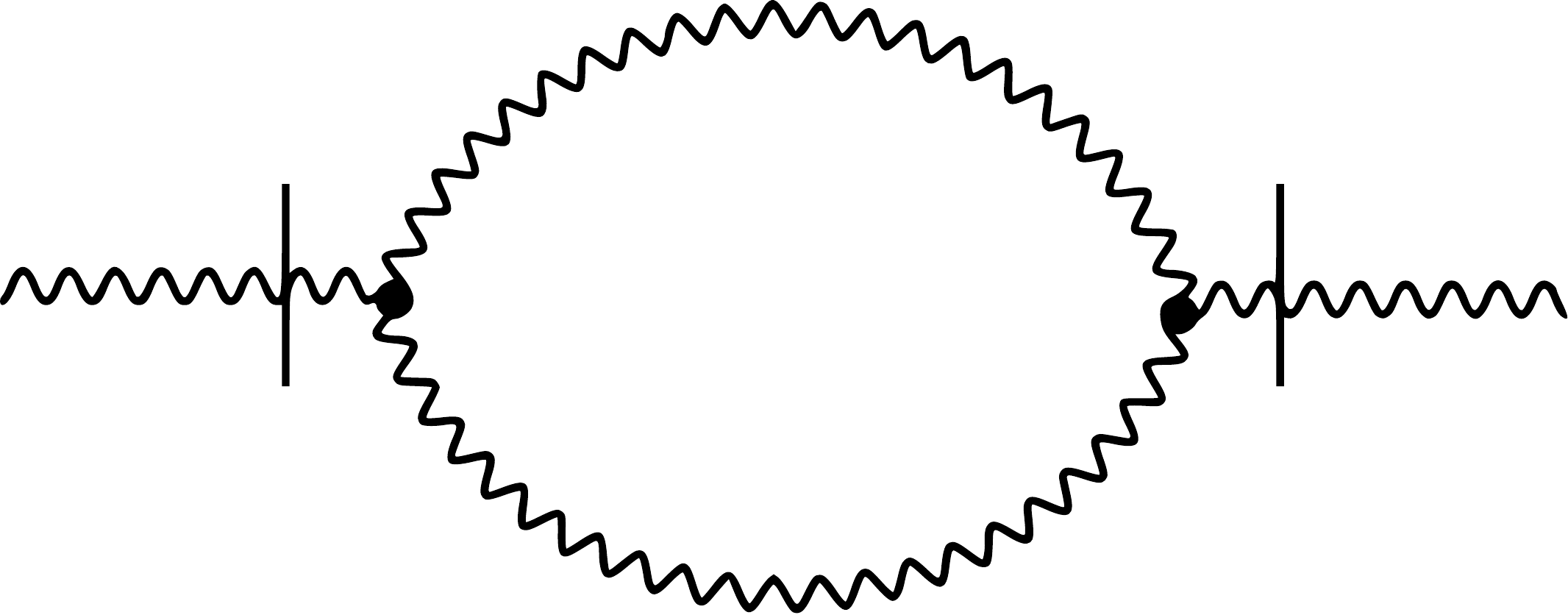}
    = 
    Z_1 + \frac{g^2}{16\pi\varepsilon}  +\dots,
\label{greenh'h'}
\\
\langle h'h \rangle_{\text{1-ir}} &=
    i\omega - \sigma q^2 Z_2 + 
    \includegraphics[height=7ex, valign=c]{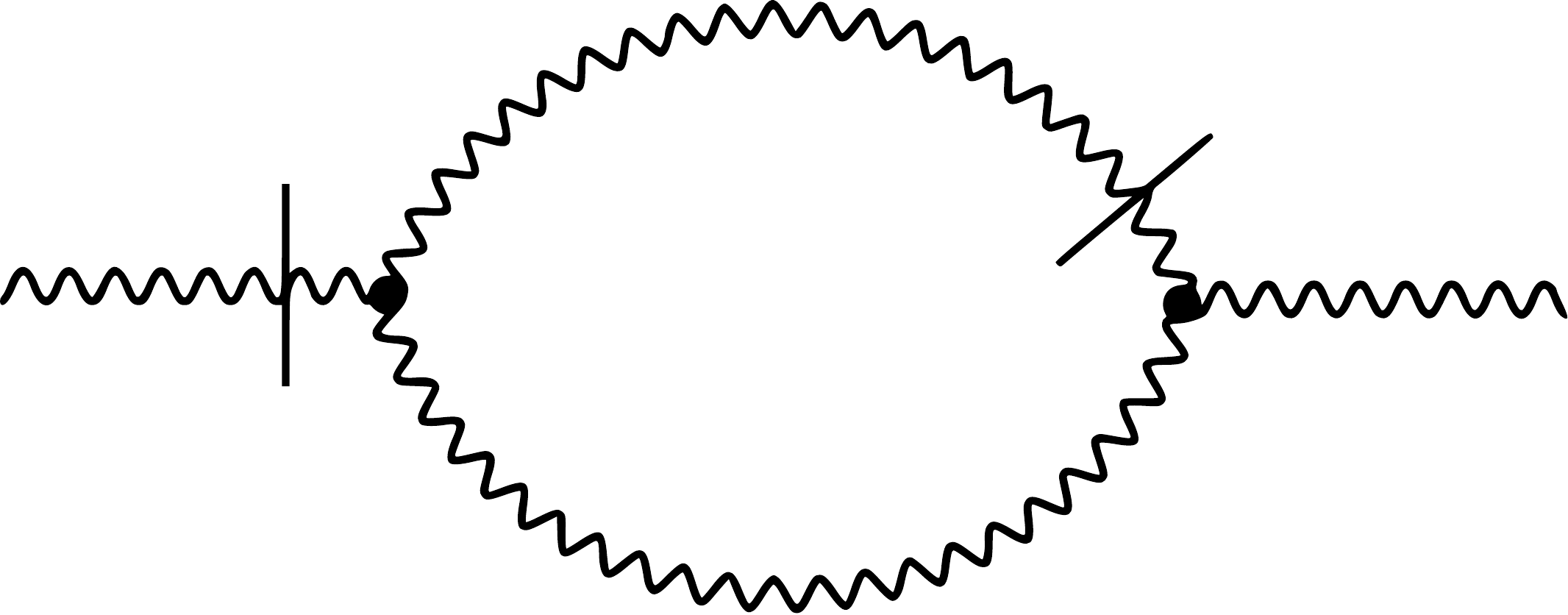} =
    i\omega - \sigma q^2 Z_2+\dots,
    \label{diaF}
    \\
    \langle \theta \theta' \rangle_{\text{1-ir}} &=
    i\omega - \sigma u q^2 Z_3 + \includegraphics[height=7ex, valign=c]{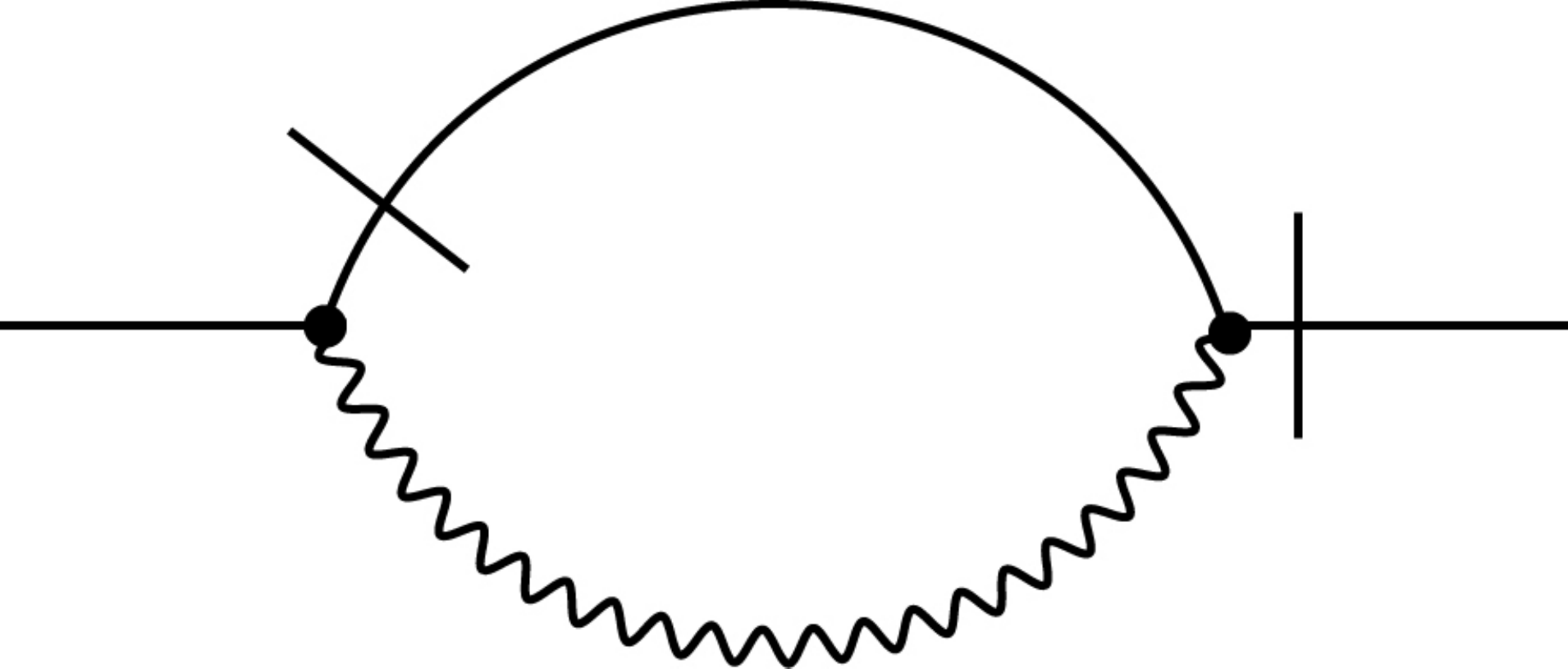} = {} \notag
    \\
    {} &= 
    i\omega - \sigma uq^2 \left[ Z_3 - \frac{w^2}{8\pi\varepsilon}\, \frac{(u-1)}{u(u+1)^2} +\dots \right],
    \label{diaX}
    \\
    \langle \theta \theta' h \rangle_{\text{1-ir}} &=
    \includegraphics[height=6ex, valign=c]{vertexthetatheta_htwisted.pdf} +
    \includegraphics[height=9ex, valign=c]{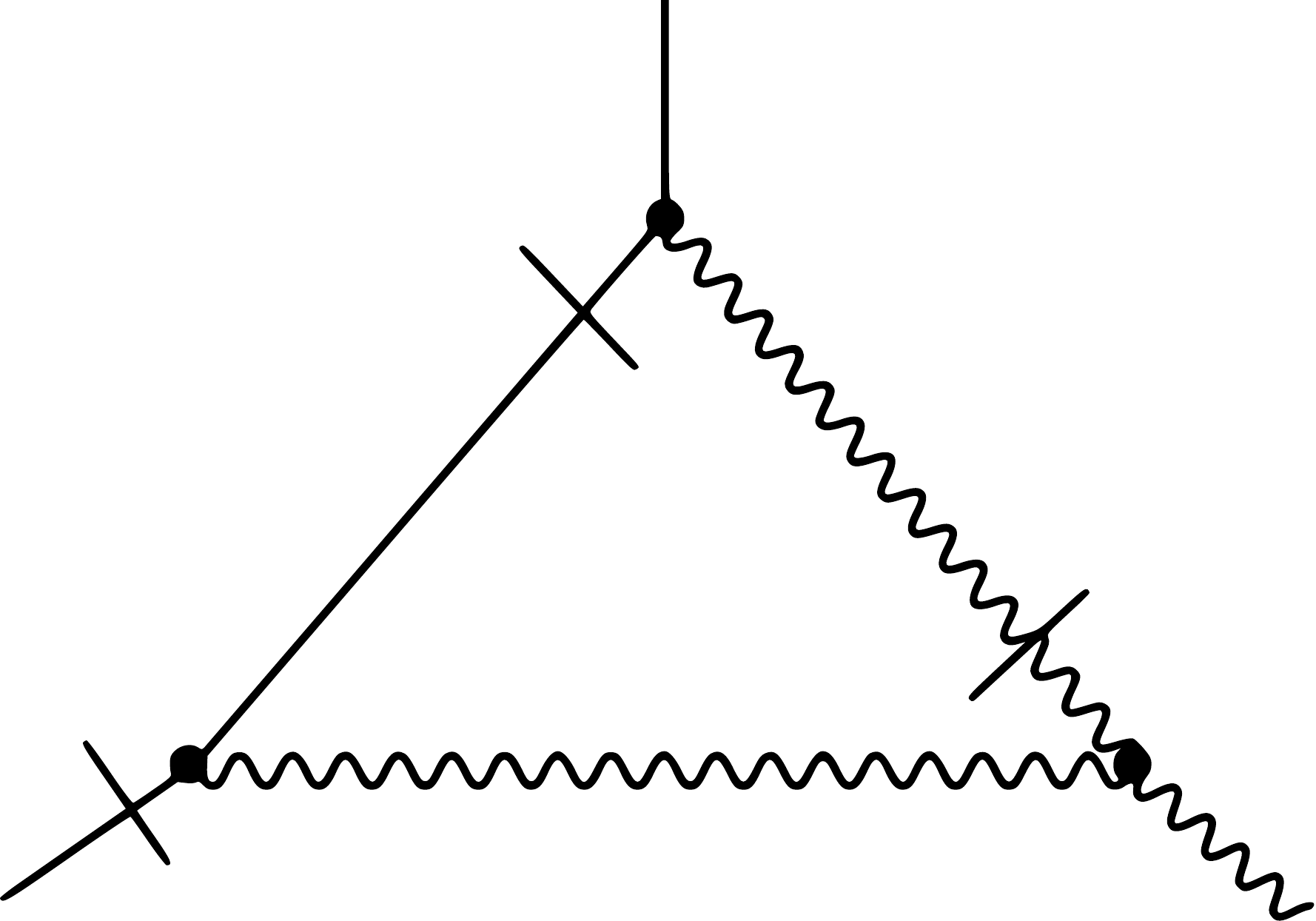} +
    \includegraphics[height=9ex, valign=c]{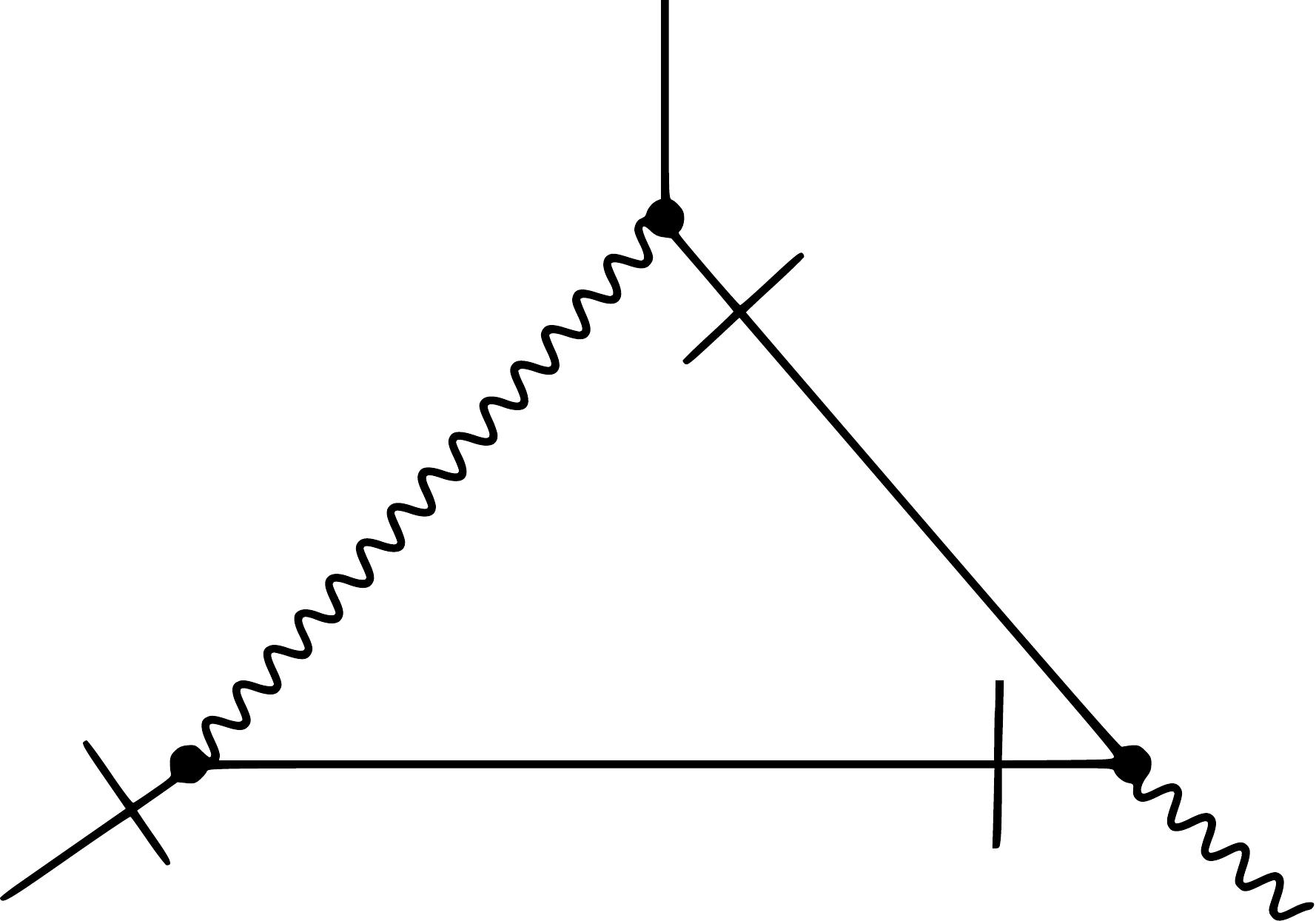} +
    \includegraphics[height=9ex, valign=c]{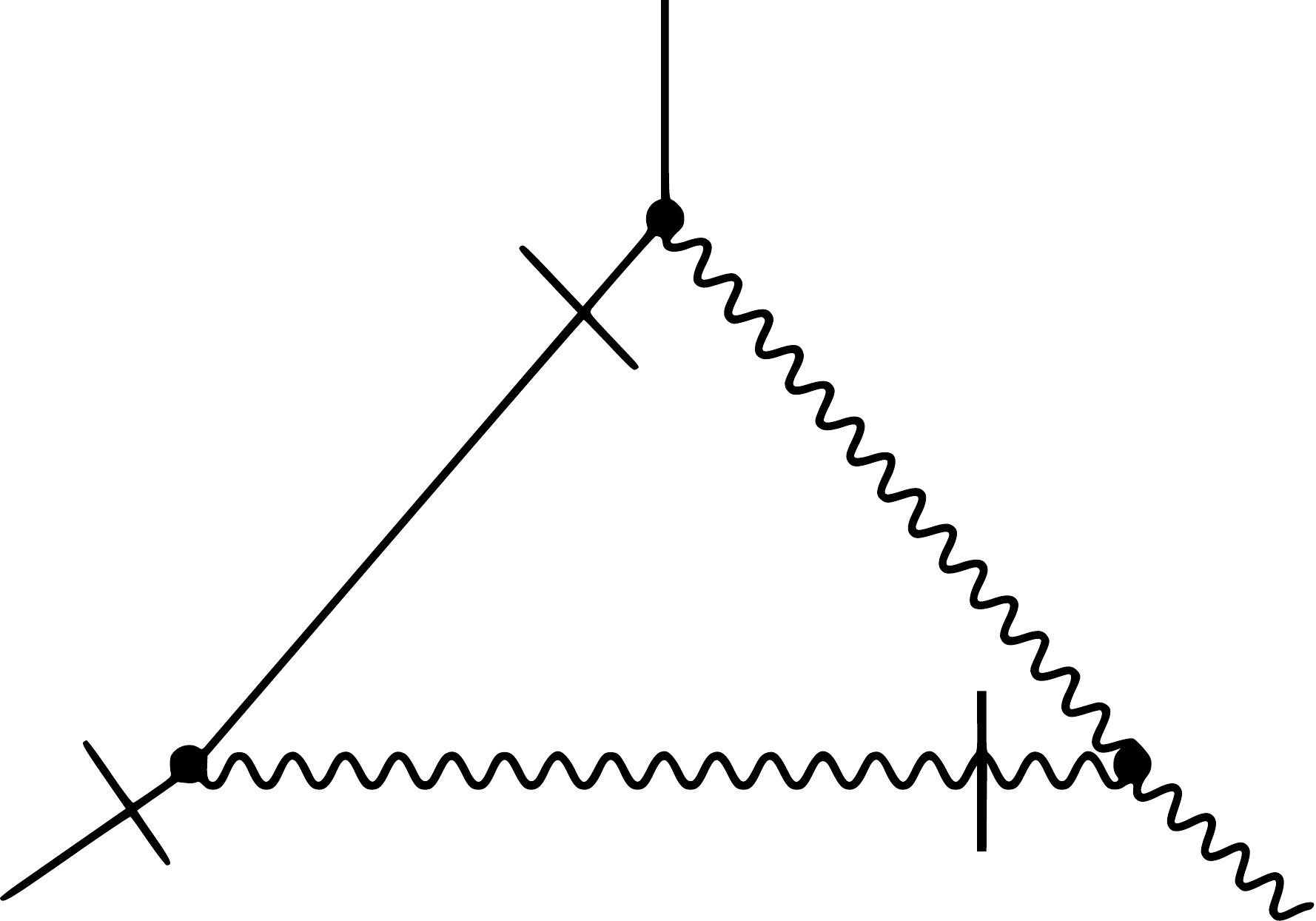}  = {} 
    \notag
    \\ {} &=
    w\sigma^{{3}/{2}} \,({\bf p\,q}) \left[ Z_4 - \frac{1}{8\pi\varepsilon} \frac{w(w-g)}{(u+1)^2}+\dots\right]. 
    \label{dialast}
\end{align}
Here the ellipses stand for UV finite parts 
(that is,  
finite at $\varepsilon \to 0$); note that the diagram in~(\ref{diaF}) has no divergent part.
We expressed the bare parameters in terms of their renormalized analogs using the relations~(\ref{renP}), (\ref{renC}) and definitions of coupling constants. In the one-loop accuracy, we keep the renormalization constants only in the tree-like (loopless) terms, while in the diagrams they are set to unities.

The renormalization constants $Z_i$ absorb all poles in $\varepsilon$, and therefore they can be found from the requirement that the Green's functions $\Gamma_R$ be UV finite
(that is, have no poles in $\varepsilon$).

The diagrams correspond to integrals, for the calculation of which it is convenient to use the following expressions:
\begin{align}
    &\int \frac{d\omega}{(2\pi)}\,
    \frac{1}{(-i\omega+a)(\omega^2+b^2)} = \frac{1}{2b(a+b)},
    \label{freq2}
    \\
    &\int\frac{d\omega}{(2\pi)}\,\frac{1}{(\omega^2 + a^2)(\omega^2 + b^2)} = \frac{1}{2ab(a+b)},
    \label{freq1}
    \\
    &\int \frac{d\omega}{(2\pi)}\,\frac{1}{(\omega^2 + a^2)(-i\omega + b)(-i\omega + c)} = \frac{1}{2a(a+b)(a+c)},
    \label{freq3}
    \\
    &\int \frac{d\omega}{(2\pi)}\,\frac{1}{(-i\omega + a)^n(i\omega + a)^l} = 
    \frac{(n+l-2)!}{(n-1)!\,(l-1)!}\, \frac{1}{(2a)^{n+l-1}},
    \label{freq4}
\end{align}
that can be obtained via residue integration\footnote{In the last expression it is assumed that~$n \geq 1$, $l \geq 1$ and $0! = 1$.}.
Then, for any vector~${\bf k}$ one has
\begin{equation}
    \label{mod}
    \int d{\bf k}\,f(k) \frac{k_i\,k_j}{k^2}
    =
    \frac{\delta_{ij}}{d}\,
    \int d{\bf k}\,f(k), \qquad k=|{\bf k}|
\end{equation}
with
\begin{equation}
    \label{mod1}
        \int d{\bf k}\,f(k) = \int d\Omega \int_0^{\infty}\, dk k^{d-1}\, f(k)=
        S_d \int_0^{\infty}\, dk\, k^{d-1}\, f(k),
\end{equation}
where the integral over the angles is:
\begin{equation}
    \label{angleint}
    \int d\Omega = S_d = \frac{2 \pi^{d/2}}{\Gamma(d/2)},
\end{equation}
$S_d$~being the surface area of a unit sphere in the $d$-dimensional space.

Let us begin with the diagram %~\includegraphics[height=5ex, valign=c]{h'h'.pdf} 
in equation~(\ref{greenh'h'}). According to the definitions given earlier, the corresponding integral is written in the form:
\begin{equation}
g^2\sigma^3\mu^{\varepsilon}
    \int \frac{d\omega}{(2\pi)}\, \frac{d{\bf k}} {(2\pi)^{d}}
     \,
    \frac{ ({\bf q} + {\bf k}, -{\bf k})({\bf k},-{\bf q}-{\bf k})}{(\omega^2 + \sigma^2({\bf q}+{\bf k})^4)\,(\omega^2 + \sigma^2 k^4)}.
    \label{PAP}
    \end{equation}
Integration over the frequency using~(\ref{freq1}) gives:
\begin{equation}
\frac{g^2\mu^{\varepsilon}}{2}
    \int \frac{d{\bf k}}{(2\pi)^{d}}\,
    \frac{( {\bf q}{\bf k}+ k^2)^2}
    {k^2 ({\bf q}+{\bf k})^2 (k^2 + ({\bf q}+{\bf k})^2) }.
    \label{Kvasimodo}
\end{equation} 

Let us briefly address the IR regularization. In the original diagrams~(\ref{greenh'h'})~-- (\ref{dialast})
and the corresponding integrals like~(\ref{PAP}), (\ref{Kvasimodo}),
it is provided by the external momenta like ${\bf q}$ and  ${\bf p}$. However, here we use the simplest way of calculation of UV divergent parts, 
where the integrands are expanded in the external momenta up to desired orders. Then, a certain artificial IR cut-off must be introduced.
According to the general statement, renormalization constants in the MS scheme do not depend on the specific choice of the IR regularization; see, e.g. section~3.19  in~\cite{Vasiliev}. For convenience, we use the simplest sharp cut-off of integration momenta at a certain value $k=m>0$.

For the Green's function $\langle h'h' \rangle_{\text{1-ir}}$ the divergence index is $\delta_{\Gamma}=0$ (see section~3), so the corresponding diagram is logarithmically divergent and can be calculated directly at zero external momentum ${\bf q}=0$. Then the integral in~(\ref{Kvasimodo}) turns to:
\begin{equation}
 \frac{g^2\mu^{\varepsilon}}{4}   \int \frac{d{\bf k}}{(2\pi)^d}\,\frac{1}{k^2} = 
\frac{g^2\mu^{\varepsilon}S_d }{4(2\pi)^d}\,
\int_m^{\infty} \, \frac{dk}{k^{1+\varepsilon}} =
     \frac{g^2 S_d }{4(2\pi)^d}\, \left(\frac{\mu}{m}\right)^{\varepsilon}\,
    \frac{1}{\varepsilon} \
= \frac{g^2}{8\pi\varepsilon}\, + O(\varepsilon^0),
\end{equation}
where the integration is easily performed using equations~(\ref{mod1}) and~\ref{angleint}), with the IR cut-off at $k=m$ in~(\ref{mod1}).
Taking into account the symmetry coefficient $1/2$ in the diagram,
we arrive at the result presented in~(\ref{greenh'h'}).

To eliminate the pole in $\varepsilon$ in~(\ref{greenh'h'}), the renormalization constant~$Z_1$ in~the MS scheme must be chosen in the form presented in~(\ref{Zs}). 
We recall that the one-loop approximation for $Z_1^{-1}$ coincides with the exact result; see the remark below equation~(\ref{Zs}) and papers~\cite{Wiese2}~-- \cite{Wiese}.

The other three Green's functions~(\ref{diaF})~-- (\ref{dialast}) are quadratically divergent,  $\delta_{\Gamma}=2$ (see section~3), so we should 
retain in the corresponding integrals the second-order terms in the expansion in external momenta: ${q}^2$ for the two-point functions~(\ref{diaF}), (\ref{diaX}) and $({\bf pq})$ for the three-point function~(\ref{dialast}). The coefficients are calculated in a similar way with the 
aid of the expressions~(\ref{freq2})~-- (\ref{angleint});
the resulting renormalization constants were given in~(\ref{102}) and~(\ref{Zs}).

The calculation shows that, due to a certain cancellation, the diagram in~(\ref{diaF}) has no pole in $\varepsilon$ in agreement with the exact result $Z_2=1$ for the KPZ model~\cite{Wiese2}~-- \cite{Wiese}.

Also note that the sum of the three diagrams in equation~(\ref{dialast}) reduces to a  simpler expression:
     \begin{equation}      
       \frac{-w\sigma^{{3}/{2}} \,({\bf p\,q})}{4\pi\varepsilon} \left\{ \frac{w^2}{2(1+u)^2} - \frac{g w(u+3)}{4(u+1)^2} + \frac{g w}{4(u+1)} \right\} =
     \frac{-w\sigma^{{3}/{2}} \,({\bf p\,q})}{8\pi\varepsilon} \frac{w(w-g)}{(u+1)^2}. 
  \label{Triple}       
\end{equation}
It vanishes for $g=w$ in agreement with the general fact that the Galilean symmetry~(\ref{Galileo})~-- (\ref{galileo9}) for the full model~(\ref{Action}) emerges at $g=w$ (or equivalently at $\lambda=\alpha$), and the renormalization constant $Z_4$ 
in~(\ref{dialast}) turns to $Z_4=1$ identically; see the discussions in the very end of sections~\ref{Model}~and~\ref{QFT}.

\end{document}